\documentclass[10pt,letterpaper]{article}
\usepackage[top=0.85in,left=0.85in,footskip=0.75in,marginparwidth=2in]{geometry}

\usepackage[utf8]{inputenc}

\usepackage{cite}

\usepackage{nameref,hyperref}

\usepackage{microtype}
\DisableLigatures[f]{encoding = *, family = * }

\raggedright
\usepackage{changepage}

\usepackage[aboveskip=1pt,labelfont=bf,labelsep=period,singlelinecheck=off]{caption}

\makeatletter
\renewcommand{\@biblabel}[1]{\quad#1.}
\makeatother

\usepackage{lastpage,fancyhdr,graphicx}
\usepackage{epstopdf}
\fancyfootoffset[L]{2.25in}

\usepackage{color}

\definecolor{Gray}{gray}{.25}

\usepackage{graphicx}

\usepackage{sidecap}

\usepackage{wrapfig}
\usepackage[pscoord]{eso-pic}
\usepackage[fulladjust]{marginnote}
\reversemarginpar

\begin{document}

\begin{flushleft}
{\Large
\textbf\newline{Feasibility of Concurrent $^{1}$H MRS \& $^{31}$P MRSI  at 7T: Brain Energy Metabolism Responses to Hyperglycemia}
}
\newline
\\
Mark Widmaier\textsuperscript{1,*},
Brooke C. Matson\textsuperscript{2},
Uzay Emir\textsuperscript{1,3,+},
Janice J. Hwang\textsuperscript{2,4,+},

\bigskip

\bf{1} Department of Radiology, University of North Carolina, North Carolina, USA
\\
\bf{2} Department of Medicine, Division of Endocrinology and Metabolism, University of North Carolina, North Carolina, USA
\\
\bf{3} Lampe Joint Department of Biomedical Engineering and the Biomedical Research Imaging Center, University of North Carolina, North Carolina, USA
\\
\bf{4} Department of Cell Biology and Physiology, University of North Carolina, North Carolina, USA

\bigskip

* mark\_widmaier@med.unc.edu\\
\bf{+} Joint co-senior ship
\end{flushleft}

\section*{Abstract}
How the human brain adjusts fuel handling and its bioenergetic state during changing glucose levels remains difficult to assess noninvasively. In this study, we established an interleaved multinuclear 7~T MR spectroscopy protocol to track a glucose-related $^{1}$H signal alongside $^{31}$P measures of high-energy phosphate metabolism during a hyperglycemic clamp.

Five healthy adults completed a morning, fasted infusion experiment consisting of baseline, ramp-up, and hyperglycemic stages over $\sim$120~min. Short-block, short-TE $^{1}$H single-voxel spectroscopy (STEAM, TE~=~11~ms; mean block duration $5.71\pm0.62$~min) was acquired in frontal cortex and quantified using the composite Glucose+Taurine (Glc+Tau) measure. $^{31}$P was acquired with rapid 3D PETALUTE MRSI using an ultrashort echo time (UTE; TE~=~65~\textmu s; 381~s per block), and high-energy phosphate ratios were derived from a posterior cortical region of interest.

Across participants, $^{1}$H Glc+Tau increased with blood glucose and showed significant elevations from baseline into hyperglycemia. In parallel, $^{31}$P ratios exhibited smaller but significant glycemia-linked responses: both PCr/Pi and $\gamma$ATP/Pi increased with blood glucose and differed across glucose clamp stages. These findings show that $^{1}$H and $^{31}$P MRS(I) can be interleaved at 7~T to measure energy metabolism within a single session.

\section{Introduction}
Cerebral energy metabolism relies on efficient glucose delivery and ATP regeneration, with oxidative phosphorylation as a major pathway and phosphocreatine (PCr) acting as a high-energy phosphate buffer \cite{Attwell2001EnergyBudget,Harris2012SynapticEnergy,Wyss2000CreatineMetabolism,Wallimann2011CKSystem}. Given the brain's high energetic demand across the human life course \cite{Pontzer2021Science,Kuzawa2014PNAS,Aronoff2022IJO,Kuzawa2019PNAS}, disruptions in energy supply or utilization are implicated in diverse diseases, including neurodegeneration, with well-described regional reductions in cerebral glucose utilization in Alzheimer's disease\cite{Mosconi2013NatRevNeurol} and cancer, where widespread metabolic reprogramming supports cancer growth and survival \cite{VanderHeiden2009Science,Pavlova2016CellMetab}. In parallel, rising rates of childhood and adult obesity worldwide have sharpened interest in central mechanisms that influence appetite, glycemic control, and long-term neurocognitive health \cite{Caprio2020NatMetab,Schwartz2005Science}. Neuroimaging findings supports association in children and adolescents  between excess adiposity and altered brain structure/connectivity in cognitive-control and reward-related networks \cite{Kaltenhauser2023JAMANO,Laurent2020JAMAPeds,Rapuano2020PNAS}. The same trend is shown in adults, where diabetes and obesity are associated with altered brain structure \cite{Dekkers2019Radiology} as well as  brain activity \cite{Li2023MolPsychiatry,Tataranni1999PNAS,Page2011JCI} and brain metabolism \cite{Volkow2009Obesity,Hwang2017JCIInsight}.
These alterations are clinically relevant, as type~2 diabetes is associated with increased risk of cerebrovascular events and microvascular brain complications, including cognitive dysfunction and mood/behavioral changes \cite{vanSloten2020LancetDiabetesEndo}.However, the metabolic underpinnings of these changes in brain structure and function remain incompletely understood. While preclinical animal models have provided important insights into the cellular mechanisms of these observations, they are insufficient for modeling the high-level structural and functional architecture of the human central nervous system. Therefore, to resolve the precise impact of metabolic dysregulation on neural integrity, translational frameworks incorporating multimodal human neuroimaging of metabolic flux are critical.

Glucose is the dominant fuel of the human brain, but not all glucose taken up is immediately oxidized. PET and MR-based work shows that, even at rest, substantial fractions of cerebral glucose metabolism proceed through non-oxidative pathways often summarized as aerobic glycolysis \cite{Vaishnavi2010PNAS,Goyal2014CellMetab,Blazey2018PLOSOne}. This balance between oxidative and non-oxidative processing varies across brain regions and is thought to support ongoing signaling and cellular maintenance demands \cite{Goyal2014CellMetab,Harris2012Neuron}. In metabolic disease, cerebral glucose handling appears to be altered: studies that acutely increase blood glucose under controlled conditions report attenuated intracerebral glucose responses in obesity and type~2 diabetes, with evidence that these abnormalities can improve with better glycemic control \cite{Gruetter1996JCBFM,Seaquist2005Metabolism,Hwang2017JCIInsight,SanchezRangel2022Diabetologia}.
A recent meta-analysis further indicates that insulin resistance modulates brain glucose metabolism in a state-dependent manner (fasting vs.\ hyperinsulinemia) \cite{Jensen2025JCEM}, and mechanistic work suggests reduced cerebral glucose transport capacity even in younger adults with obesity \cite{Gunawan2024Obesity}.

Magnetic resonance spectroscopy (MRS) provides complementary, non-ionizing access to these processes. At ultra-high field (7~T), $^{1}$H MRS benefits from increased SNR and improved spectral dispersion, and has been implemented under controlled glycemic conditions even in challenging regions such as the hypothalamus and prefrontal cortex \cite{Moheet2014MRM,Park2025NBM}. Cerebral glucose and related metrics can then be quantified using glucose-specific strategies (e.g., J-difference editing of the $\beta$-glucose resonance) or direct fitting of the H1-$\alpha$-glucose peak, as well as more robust composite readouts suited to repeated short blocks \cite{Kaiser2016MRM,Kuribayashi2024JMRI}. In parallel, $^{31}$P MRS probes high-energy phosphates (PCr, ATP) and inorganic phosphate (Pi), providing sensitive markers of energetic state and phosphorylation potential \cite{Rietzler2022JNeuroradiol}. However, relatively few studies have interrogated cerebral $^{31}$P energetics during acute glycemic manipulations, despite suggestive evidence that brain high-energy phosphate ratios can respond measurably to glycemic variation and metabolic phenotype \cite{Oltmanns2008AJPRegu,Schmoller2010JCBFM}.

A practical barrier for dynamic, intervention-based studies is acquisition efficiency: repeated $^{1}$H measurements must be balanced against the longer repetition times and lower sensitivity of multinuclear acquisitions. Interleaved heteronuclear strategies have been demonstrated for human brain $^{1}$H/$^{31}$P MRS by inserting $^{1}$H readouts within the $^{31}$P acquisition cycle, enabling combined sampling without repeated setup and shimming \cite{Chang1991MRI,Gonen1994MRM,AriasMendoza1996NBM}. Such approaches are increasingly relevant at 7~T, where ultra-high-field gains in SNR (signal-to-noise ratio) and spectral resolution can be used for time-resolved metabolic phenotyping, but also impose SAR (specific absorption rate), $B_0$-inhomogeneity, and hardware constraints (e.g., double-tuned coil trade-offs) \cite{Choi2020DoubleTunedReview}.

In this work, we develop and evaluate an interleaved $^{1}$H/$^{31}$P protocol during a  glucose infusion paradigm at 7~T, pairing repeated $^{1}$H SVS (single voxel spectroscopy) in the frontal cortex with a whole-brain $^{31}$P MRSI. We tested whether $^{1}$H MRS-measured glucose-related signal and $^{31}$P-derived energetic ratios are altered with blood glucose within the same interleaved acquisition. By combining substrate-sensitive and energetic readouts within a single session, this framework aims to provide a practical foundation for future studies of acute metabolic flexibility and energetic regulation, including in altered metabolic health such as obesity and diabetes.

\section{Methods}

\subsection{Participants and ethics}
Five healthy adults (4 male, 1 female; age $29.0 \pm 9.5$~years, range 20 - 45~years; BMI $22.21 \pm 1.43$~kg/m$^2$; HbA1c $4.94 \pm 0.37$\%) were scanned.The study was approved by the University of North Carolina at Chapel Hill (UNC) Institutional Review Board. Lean, healthy adults without diabetes were recruited from the general population. Inclusion criteria were age 18-45~years, hemoglobin A1c $<6.5\%$, and body mass index (BMI) 17-25~kg/m$^2$. Exclusion criteria included active chronic medical problems, prescription medication use, body-weight change $>5\%$ in the preceding 6~months, and tobacco use, high-risk alcohol use, or illicit drug use. All participants provided written informed consent.

\subsection{Hyperglycemic experiment procedure}
Participants arrived at UNC's Biomedical Research Imaging Center in the morning after an overnight fast. A peripheral intravenous catheter was inserted into each arm: one for infusion of 20\% dextrose in water and one for blood sampling as previously described\cite{Hwang2017JCIInsight,Gunawan2024Obesity,Hwang2018JCI_HypoglycemiaUnawareness}. Blood glucose was measured using a glucose oxidase method (YSI 2500, Yellow Springs Instruments, Yellow Springs, OH, USA). During baseline interleaved $^{1}$H MRS and $^{31}$P PETALUTE acquisitions (approximately $-50$ to $0$~min; Figure~\ref{fig:fig1}A), glucose was sampled every 10~min. After completion of baseline scans, dextrose infusion began and glucose was sampled every 5~min to guide manual adjustment of the infusion rate, targeting a stable blood glucose concentration of 180~mg/dL (hyperglycemic clamp) \cite{DeFronzo1979Clamp}.

\textbf{Glucose-clamp-stage definition:} The interleaved protocol comprised a baseline stage, a ramp-up stage after infusion start, and sustained hyperglycemia (Figure~\ref{fig:fig1}A). Baseline was defined as four interleaved blocks of $^{1}$H MRS and $^{31}$P PETALUTE (Participant~1: two baseline blocks). The infusion (ramp-up) stage was defined as $^{1}$H STEAM blocks acquired without interleaved PETALUTE. This $^{1}$H-only sampling during ramp-up increased temporal resolution of the glucose-related readout while infusion rates were being adjusted, and interleaved $^{1}$H/$^{31}$P acquisitions were resumed once blood glucose was stabilized at the target hyperglycemic level. Hyperglycemia~I and Hyperglycemia~II were then defined as two interleaved $^{1}$H/$^{31}$P blocks each during sustained hyperglycemia; Participant~4 did not complete Hyperglycemia~II due to intravenous catheter difficulties. Participant-specific infusion profiles and glucose time courses are shown in Figure~2.

\subsubsection{MR system, shimming, acquisition, and reconstruction}
All scans were performed on a 7~T Siemens Magnetom (VB 17A) system using a single-channel dual-tuned $^{1}$H/$^{31}$P head coil (QED). For $^{31}$P PETALUTE, the vendor-provided global shim was used. For $^{1}$H MRS, localized shimming was performed with FASTMAP\cite{Gruetter1993FASTMAP}, and VAPOR water suppression\cite{Tkac1999VAPOR} was calibrated before baseline acquisition. Prior to Baseline 1H PETALUTE\cite{Shen2023RosetteUTE} and  MP2RAGE\cite{Marques2010MP2RAGE} anatomical images were acquired. 

\textbf{$^{1}$H MRS acquisition and voxel placement:} $^{1}$H MRS was acquired from an 8~mL voxel (20$\times$20$\times$20~mm$^3$) positioned in the frontal cortex (Figure~\ref{fig:fig1}B, blue) based on MP2RAGE\cite{Marques2010MP2RAGE} guidance. Spectra were acquired using STEAM (TE~=~11~ms) with VAPOR water suppression and 64 transients. Due to SAR constraints, the repetition time was adapted across participants, resulting in subject-specific $^{1}$H block durations (mean $5$~min $43$~sec $\pm$ 37~sec)\cite{Emir2012NBM_B1Shimming}.

\textbf{$^{31}$P PETALUTE MRSI acquisition, reconstruction, and region-of-interest (ROI) spectra:} $^{31}$P data were acquired using a 3D PETALUTE whole-brain MRSI sequence \cite{Shen2023RosetteUTE,Alcicek2025PETALUTE,Bozymski2025SciRep_PETALUTE} (361 petals, TE~=~65~\textmu s, TR~=~0.35~s, flip angle~=~25$^\circ$, spectral bandwidth~=~4.166~kHz, 256 complex time-domain points). The acquisition was prescribed at scanner isocenter with an isotropic field-of-view (FOV) of 480$\times$480$\times$480~mm$^3$ and reconstructed on a $24\times24\times24$ matrix (20~mm isotropic nominal resolution). Each PETALUTE block lasted 6~min 21~sec and consisted of three successive acquisitions, that were averaged prior to reconstruction; each acquisition had a four-time undersampled k-space readout.

\subsection{Spectral quantification}
\textbf{$^{1}$H spectral preprocessing and quantification:} $^{1}$H spectra were processed in MRSpecLab\cite{Xiao2025MRSpecLAB,MRSpecLAB_GitHub} using the standard pipeline comprising frequency/phase alignment across transients, eddy-current correction using an unsuppressed water reference, signal averaging, and residual water removal using an HLSVD-based notch filter. Quantification was performed via LCModel (version 6.3-1N)\cite{Provencher1993LCModel,Provencher2001LCModel} invoked through MRSpecLab using default settings and a study-specific basis set\cite{Emir2012NBM_B1Shimming}. Spectra were fit over 4.2 to 0.5~ppm with a spline baseline (knot spacing 0.25~ppm). Water referencing was enabled and metabolite estimates were scaled by fixing total creatine (Cr+PCr) to 8~mM, with time courses further normalized to each participant’s mean baseline value. For metabolites which showed strong anti-correlation in the fit - correlation coefficient $<-0.5$ -  within a given region, their sum was reported to improve robustness (e.g., Cr+PCr, GPC+PC, and Glc+Tau). Macromolecular contributions were modeled using an in vivo macromolecule spectrum acquired from the occipital cortex of five volunteers with an inversion-recovery sequence (TR~=~3~s, TE~=~11~ms, TI~=~0.685~s)\cite{Emir2012NBM_B1Shimming}; this macromolecule spectrum was included in the model basis and parameterized as individual macromolecule components during fitting.

\textbf{$^{31}$P PETALUTE  spectra:} Non-Cartesian reconstruction was performed offline using NUFFT with trajectory-based density compensation \cite{Fessler2003NUFFT,Pipe1999DensityComp}. A spatiotemporal  reconstruction was formulated with total generalized variation\cite{Bredies2010TGV}. The optimization problem was solved with a first-order primal-dual algorithm using fixed regularization settings across participants and blocks \cite{Chambolle2011PrimalDual}. Reconstruction yielded complex spectra per voxel, which were subsequently phase-corrected and frequency-centered prior to ROI combination. 

For quantification, voxel spectra were combined into a posterior ROI (60$\times$60$\times$40~mm$^3$, 144~mL) shifted 7~cm posteriorly along the anterior - posterior axis (Figure \ref{fig:fig1}B, orange). ROI spectra were generated by an overlap-weighted sum of voxel spectra, where weights were computed analytically from the geometric overlap between the cuboid ROI and the 20~mm isotropic voxel grid. Residual nuisance components due to aliasing  were attenuated using an HLSVD-based filter\cite{Barkhuijsen1987HSVD,Pijnappel1992HLSVD,Larsen1998DAIMI},  and a first-order phase correction accounting for TE was applied in the frequency domain\cite{Alcicek2025PETALUTE}.
The final complex spectra from the PETALUTE reconstruction pipeline were imported into MRSpecLab and quantified  with LCModel, without additional preprocessing. LCModel fitting used a study-specific $^{31}$P basis set and default regularization for baseline and linewidth estimation. Spectra were fit over a broad $^{31}$P range (5.5 to -16.9~ppm) to jointly model PCr, Pi, and ATP resonances. For analysis and visualization, $^{31}$P measures were expressed relative to Pi and further normalized to each participant's mean baseline ratio.

\subsection{Statistical analysis}
Associations between blood glucose and MRS-derived measures (Glucose+Taurine (Glc+Tau); PCr/Pi; $\gamma$ATP/Pi) were assessed using linear mixed-effects models with a random intercept per subject, fit by maximum likelihood. Within-subject associations were additionally assessed using repeated-measures correlation.
Clamp-stage differences (Baseline, Ramp-up, Hyperglycemia~I, Hyperglycemia~II, and pooled Hyperglycemia~(I+II) where applicable) were evaluated using mixed-model marginal tests and post-hoc comparisons consistent with the annotations in Figure~\ref{fig:fig4}.

\section{Results}

\subsection{Glucose infusion protocol and representative spectra}
\begin{figure}
\centering
\includegraphics[width=\textwidth]{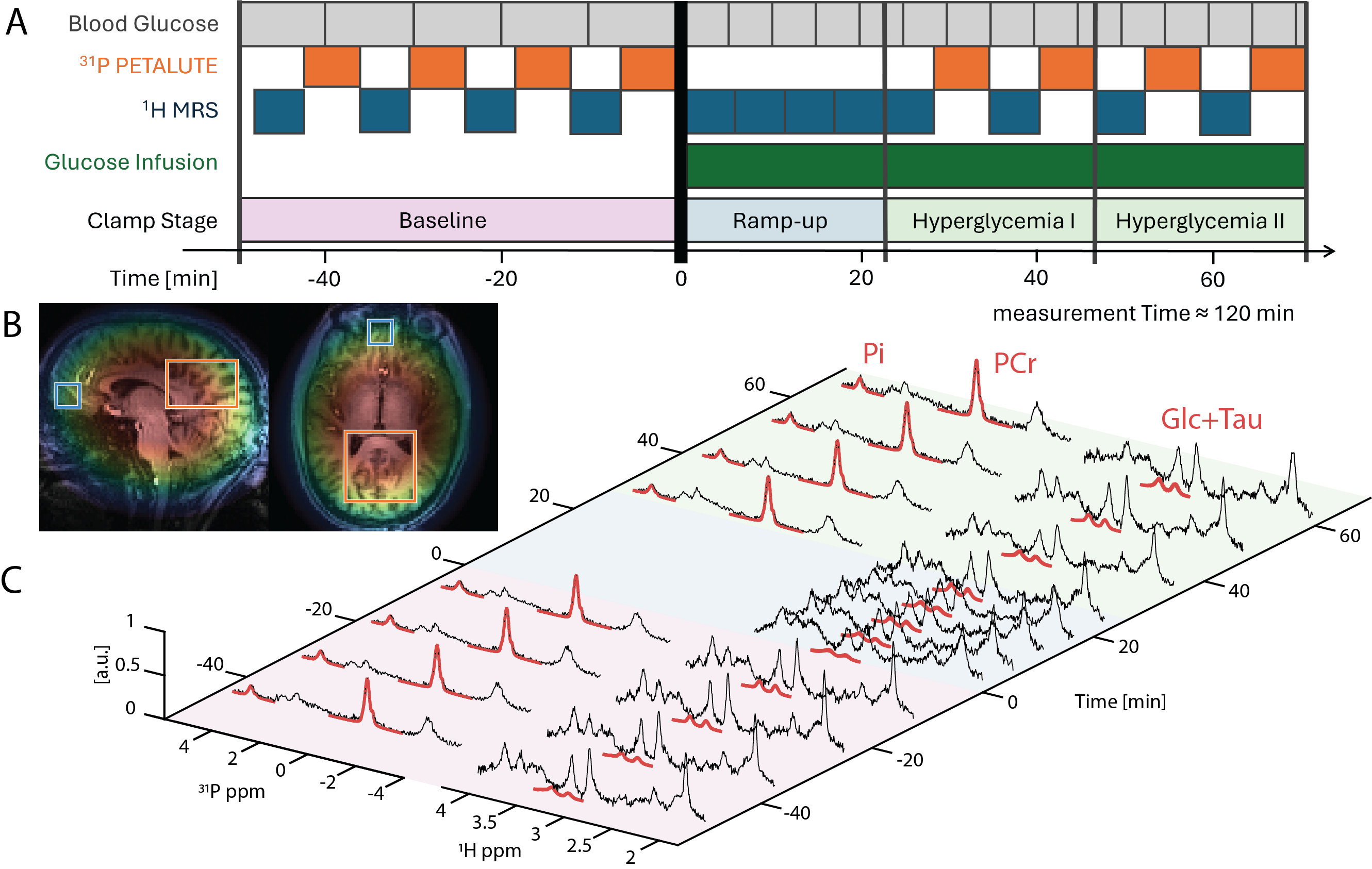}
\caption{\textbf{Interleaved $^{1}$H/$^{31}$P MRS protocol during hyperglycemic clamp at 7~T.}
(\textbf{A}) Timing diagram showing repeated blood-glucose sampling (gray), interleaved $^{31}$P PETALUTE MRSI blocks (orange) and $^{1}$H MRS blocks (blue) across baseline, ramp-up, and two hyperglycemic stages. The glucose infusion period is indicated in green; total measurement time $\approx$120~min.
(\textbf{B}) Anatomical reference (MP2RAGE) overlaid and $^{31}$P PETALUTE first time point image with ROIs marked (Participant~3): frontal $^{1}$H SVS voxel (20$\times$20$\times$20~mm$^{3}$, blue) and the posterior $^{31}$P used for PETALUTE voxel averaging (60$\times$60$\times$40~mm$^{3}$, orange).
(\textbf{C}) Representative time-resolved spectra (Participant~3) acquired throughout the session for $^{31}$P (left; Pi and PCr highlighted) and $^{1}$H (right; Glc+Tau region highlighted).
\label{fig:fig1}}
\end{figure}

\begin{figure}
\center
\includegraphics[width=0.95\textwidth]{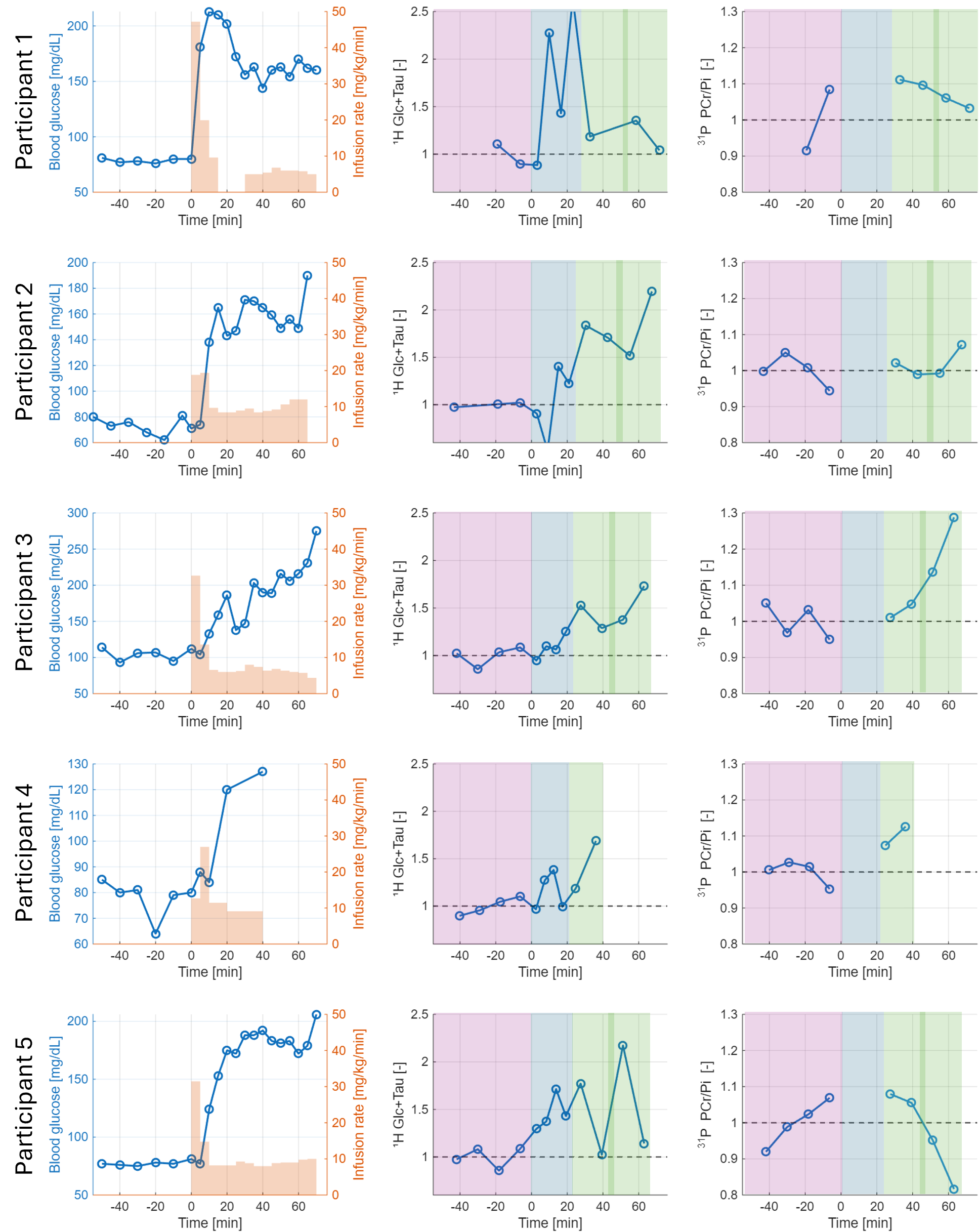}
\caption{\textbf{Participant-specific clamp performance and metabolite time courses.}
Rows show individual participants. Left column: blood glucose (blue, left axis) and glucose infusion rate (orange, right axis) over time. Middle column: baseline-normalized $^{1}$H Glc+Tau measured from the frontal SVS voxel. Right column: baseline-normalized $^{31}$P PCr/Pi derived from PETALUTE ROI spectra. Shaded backgrounds indicate clamp-stages (baseline, ramp-up, Hyperglycemia~I, Hyperglycemia~II) using the same color scheme as Figure~\ref{fig:fig1}. Dashed horizontal lines denote the baseline normalization reference (value = 1).
\label{fig:fig2}}
\end{figure}

ROI positioning is shown on the MP2RAGE anatomical reference overlaid with the first $^{31}$P PETALUTE time point, marking the frontal $^{1}$H SVS voxel (20$\times$20$\times$20~mm$^{3}$, blue) and the posterior ROI used for $^{31}$P voxel averaging (60$\times$60$\times$40~mm$^{3}$, orange) (Figure~\ref{fig:fig1}B). The PETALUTE intensity map further highlights the characteristic sensitivity profile of the single-channel volume coil, with highest $^{31}$P SNR near the brain center and reduced sensitivity toward peripheral/anterior regions, motivating the posterior ROI selection for robust $^{31}$P quantification. Representative ROI placement and time-resolved spectra illustrate stable, well-resolved $^{31}$P resonances (including PCr and Pi) alongside the expected increase in the $^{1}$H Glc+Tau region during acute hyperglycemia (Figure~\ref{fig:fig1}C).

Figure~\ref{fig:fig2} summarizes individual glucose clamp performance and metabolic readouts. Blood glucose increased from baseline values to sustained hyperglycemic levels during infusion. In parallel, the normalized $^{1}$H Glc+Tau signal increased above baseline (dashed line at 1), with peak responses ranging from $\sim$1.7 to $\sim$2.5 across participants and a generally sustained elevation during hyperglycemia. The $^{31}$P PCr/Pi ratio remained close to unity for most participants, with inter-individual divergence: Participant~3 showed a progressive increase (up to $\sim$1.3), whereas Participant~5 exhibited a late decrease (down to $\sim$0.8) during the second hyperglycemic stage (Figure~\ref{fig:fig2}). For visualization of temporal co-variation across modalities on a common scale, the corresponding $z$-scored trajectories are provided in Supplementary Figure~S1.

\subsection{Coupling between blood glucose, $^{1}$H Glc+Tau, and high-energy phosphate ratios}
\begin{figure}
\center
\includegraphics[width=\textwidth]{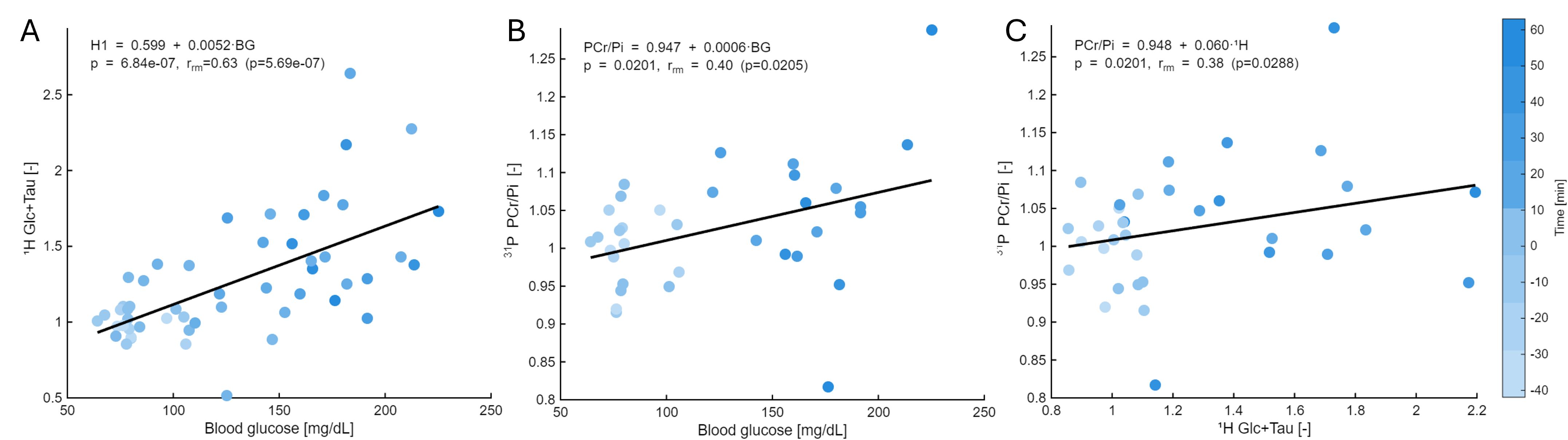}
\caption{\textbf{Associations between blood glucose, $^{1}$H Glc+Tau, and $^{31}$P PCr/Pi.}
(\textbf{A}) $^{1}$H Glc+Tau versus blood glucose across all participants and time points. The solid line indicates the linear mixed-effects fit (random intercept per subject); the repeated-measures correlation coefficient ($r_m$) and $p$-values are reported in-panel.
(\textbf{B}) $^{31}$P PCr/Pi versus $^{1}$H Glc+Tau with linear fit and repeated-measures correlation.
(\textbf{C}) $^{31}$P PCr/Pi versus blood glucose with linear fit and repeated-measures correlation. Marker color encodes time relative to infusion start (minutes; color bar).\label{fig:fig3}}
\end{figure}

Across all sessions, the normalized $^{1}$H Glc+Tau signal increased with blood glucose (Figure~\ref{fig:fig3}A). A linear mixed-effects model (random intercept per subject) showed a robust positive association (slope $=5.17\times10^{-3}$ per mg/dL, 95\%~CI $[3.34,\,7.00]\times10^{-3}$; $p=6.84\times10^{-7}$; $r_m=0.63$). Including measurement time as an additional covariate did not indicate a systematic temporal drift ($p=0.506$), while the blood-glucose effect remained significant ($p=3.89\times10^{-3}$), supporting that the observed $^{1}$H changes primarily tracked clamp-driven glycemia, with changes in blood glucose, rather than time-in-scan effects.

High-energy phosphate ratios exhibited smaller but consistent glycemia-linked trends. $^{31}$P PCr/Pi increased with blood glucose (Figure~\ref{fig:fig3}B; slope $=6.34\times10^{-4}$ per mg/dL, 95\%~CI $[1.06,\,11.62]\times10^{-4}$; $p=0.020$; repeated-measures correlation $r_m=0.40$, $p=0.0205$) and also covaried with $^{1}$H Glc+Tau (Figure~\ref{fig:fig3}C; $r_m=0.38$, $p=0.0288$). However, when $^{1}$H Glc+Tau and blood glucose were entered jointly, $^{1}$H Glc+Tau did not explain additional variance in PCr/Pi ($p=0.924$), consistent with blood glucose being the primary predictor in this dataset. Interpreting model slopes in clinically intuitive units, a +50~mg/dL increase in blood glucose corresponded to a +0.26 increase in normalized $^{1}$H Glc+Tau (95\%~CI: +0.17 to +0.35). For $^{31}$P, the same +50~mg/dL increase corresponded to a +0.032 change in PCr/Pi (95\%~CI: +0.005 to +0.058) and a +0.046 change in $\gamma$ATP/Pi (95\%~CI: +0.018 to +0.074), indicating smaller but measurable shifts in high-energy phosphate ratios over the clamp range. 
Similar results were observed for $\gamma$ATP/Pi (Supplementary Figure~S2), which increased with blood glucose (slope $=9.13\times10^{-4}$ per mg/dL, 95\%~CI $[3.53,\,14.73]\times10^{-4}$; $p=0.00226$), and showed a positive association with $^{1}$H Glc+Tau consistent with the PCr/Pi findings.

\subsection{Glucose-clamp-stage-resolved changes} 
\begin{figure}
\center
\includegraphics[width=0.75\textwidth]{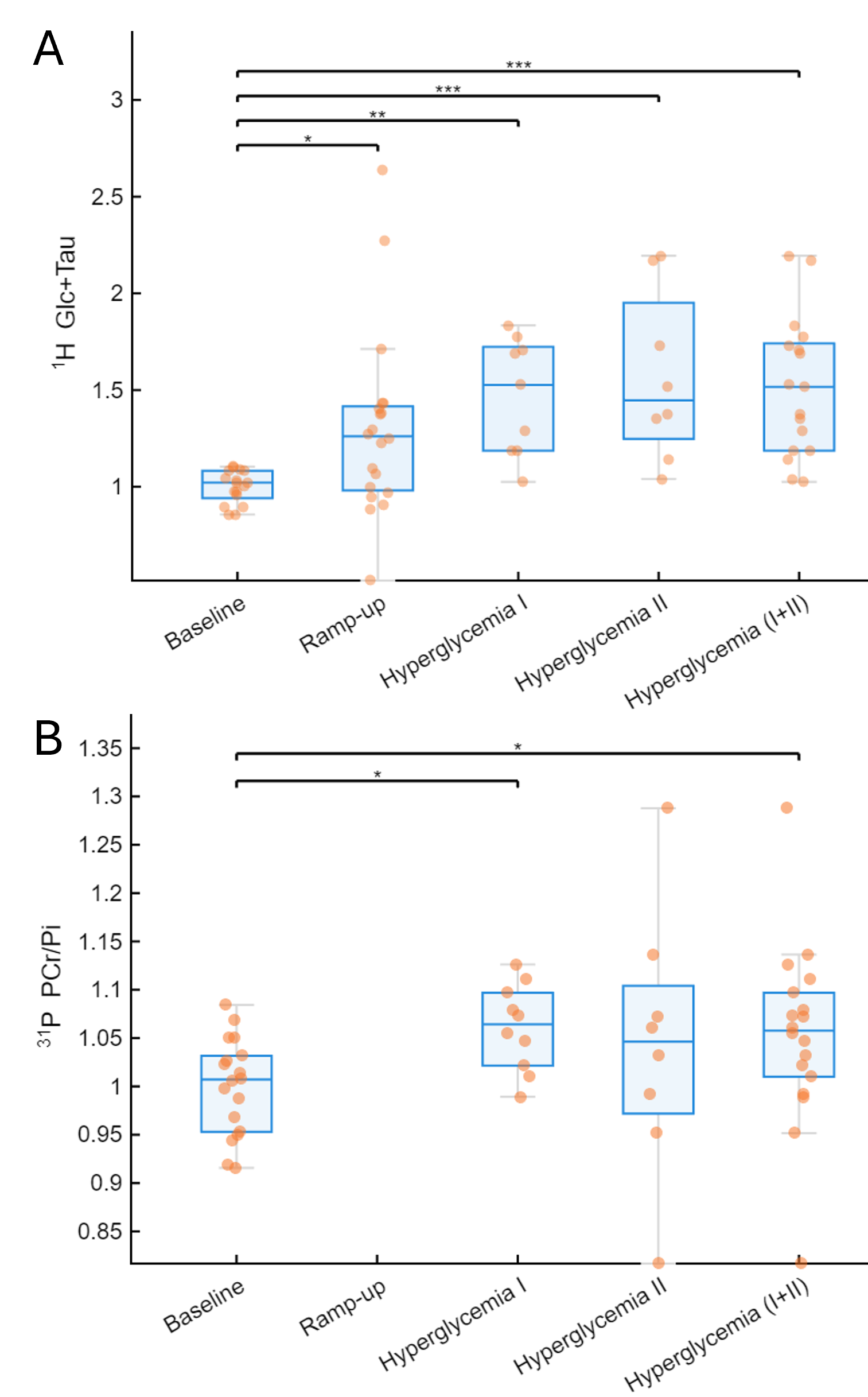}
\caption{\textbf{Clamp-stage-dependent changes in $^{1}$H Glc+Tau and $^{31}$P PCr/Pi.}
(\textbf{A}) Baseline-normalized $^{1}$H Glc+Tau across  clamp-stages (Baseline, Ramp-up, Hyperglycemia~I, Hyperglycemia~II, and pooled Hyperglycemia~(I+II)). 
(\textbf{B}) Baseline-normalized $^{31}$P PCr/Pi across the same clamp-stages.
Boxplots show median and interquartile range; whiskers indicate range and points represent individual measurements; Horizontal brackets denote significant post-hoc differences between clamp-stages ($^{*}p<0.05$, $^{**}p<0.01$, $^{***}p<0.001$).\label{fig:fig4}}
\end{figure}

Glucose-clamp-stage analysis revealed a stepwise elevation of $^{1}$H Glc+Tau from baseline to ramp-up and sustained increases during hyperglycemia (Figure~\ref{fig:fig4}A). Pairwise comparisons versus baseline showed significant increases already during ramp-up ($p<0.05$) and larger elevations during Hyperglycemia~I and Hyperglycemia~II (up to $p<0.001$; Figure~\ref{fig:fig4}A). In line with this, mixed-model marginal testing confirmed a strong clamp-stage effect on $^{1}$H Glc+Tau (clamp-stage term: $F\approx 40$, $p\le 1.5\times10^{-6}$).

For $^{31}$P, PCr/Pi was higher during hyperglycemia compared with baseline, reaching significance for Hyperglycemia~I and for the pooled hyperglycemia condition (I+II) (Figure~\ref{fig:fig4}B; $p<0.05$). While the central tendency remained elevated into Hyperglycemia~II, dispersion increased in late clamp (visible as a wider spread in Hyperglycemia~II), consistent with a higher within-subject variance in Hyperglycemia~II compared with Hyperglycemia~I (paired log-variance test: $t(3)=4.90$, $p=0.0163$).

\section{Discussion}

This study demonstrates the feasibility of an interleaved $^{1}$H/$^{31}$P MRS protocol during acute hyperglycemia generated by intravenous glucose infusion (hyperglycemic clamp) at 7~T, enabling combined tracking of $^{1}$H cerebral glucose and high-energy phosphate metabolites from $^{31}$P. Across participants and time points, $^{1}$H Glc+Tau increased with increasing blood glucose and showed clear clamp-stage-dependent elevations from baseline into hyperglycemia. In parallel, $^{31}$P ratios exhibited smaller but significant associations with blood glucose, with both PCr/Pi and $\gamma$ATP/Pi showing positive slopes versus blood glucose and clamp-stage-dependent increases. Together, these results support that acute blood glucose elevation is accompanied by measurable changes in cerebral metabolic responses,  detectable with interleaved $^{1}$H/$^{31}$P at ultra-high field.

The strong within-subject relationship between blood glucose and $^{1}$H Glc+Tau is in line with prior clamp-based MRS work demonstrating that intracerebral glucose increases with blood glucose across different brain regions in healthy individuals, and that the magnitude and kinetics of this response can be altered in metabolic disease states such as obesity and type~2 diabetes\cite{Gruetter1996JCBFM,Seaquist2005Metabolism,Heikkila2010MetabBrainDis,Hwang2017JCIInsight,SanchezRangel2022Diabetologia}. In this context, our data extend these observations to a repeated-measures, short-block protocol at 7~T with an anterior/frontal $^{1}$H voxel, where susceptibility and baseline stability constraints are more pronounced yet the Glc-related signal remained strongly coupled to blood glucose across time.

Methodologically, several 7~T studies have specifically targeted improved glucose specificity by reducing overlap with taurine/macromolecules, for example through J-difference editing of the $\beta$-glucose resonance near 3.23~ppm \cite{Kaiser2016MRM} or by fitting the H1-$\alpha$-glucose peak at 5.23~ppm \cite{Kuribayashi2024JMRI}. Compared with these more glucose-specific approaches, the composite Glc+Tau metric used here trades biochemical specificity for robustness in a time-resolved setting with SAR-driven variability in block duration, where glucose is particularly susceptible to baseline/model degeneracy. In addition, the interleaved multinuclear setup relied on a single-channel double-tuned $^{1}$H/$^{31}$P head coil, which limits SNR and constrains the use of advanced editing strategies relative to dedicated multi-channel $^{1}$H arrays typically used in glucose-edited $^{1}$H MRS at 7~T \cite{Choi2020DoubleTunedReview}.


We extended the repeated-measures $^{1}$H design by interleaving $^{31}$P measurements, enabling parallel assessment of high-energy phosphate metabolism. Interleaved $^{1}$H/$^{31}$P strategies have previously been demonstrated in the human brain in other studies \cite{Gonen1994Interleaved1H31P,AriasMendoza1996Interleaved,Chang1991Interleaved} and analogous time-sharing concepts have also been pursued for $^{1}$H/$^{13}$C protocols in the frontal lobe to monitor oxidative metabolism during labeled glucose infusion \cite{DeFeyter2018selPOCE,Jacobs2024Indirect1H13C7T,Ahmadian2025selPOCErepro,Dehghani2020Dynamic1H13Cglc,Cheshkov2017U13Cglc7T,Xiao2026Thesis}. In that sense, our work complements prior single-nucleus clamp studies by showing that dynamic $^{1}$H glucose-sensitive metrics can be acquired  while concurrently sampling $^{31}$P high-energy phosphates at 7~T.

The observed $^{31}$P effects were more modest, but directionally consistent with the limited prior literature interrogating cerebral high-energy phosphates during acute glycemic manipulations. In healthy volunteers, dextrose-induced hyperglycemia has been reported to increase cerebral phosphocreatine (PCr) and PCr/Pi over short challenges, whereas no significant behavior was detected in skeletal muscle, supporting the notion that brain high-energy phosphate pools can respond measurably even over brief glycemic perturbations \cite{Oltmanns2008Glycemic31P}. Related clamp-based work further suggests that cerebral PCr and ATP levels can vary systematically with metabolic phenotype (e.g., inverse association with BMI) under controlled clamp conditions \cite{Schmoller2010BMI31P}. Along the same line, Wardzinski et al.\ reported glucose-load-dependent modulation of high-energy phosphate ratios in normal-weight controls, with altered response patterns in obesity, highlighting that energetic readouts may vary across both glycemic level and metabolic state \cite{Wardzinski2018Metabolism}.

In our data, both PCr/Pi and $\gamma$ATP/Pi increased with blood glucose and showed clamp-stage-related shifts, supporting sensitivity of PETALUTE-derived ROI spectra to glycemia-linked energetic modulation. Notably, adding Glc+Tau to the mixed-effects model for PCr/Pi did not improve explanatory power beyond blood glucose, suggesting that, within the constraints of our design and non-identical sampling volumes, blood glucose was the dominant predictor of the observed $^{31}$P ratio changes. Physiologically, modest elevations in PCr- and ATP-related ratios may reflect altered phosphorylation potential, substrate-driven changes in oxidative metabolism, and/or shifts in Pi homeostasis; however, interpretation should remain cautious because ratios can be influenced by multiple factors such as pH, partial-volume effects and fit stability, and because $^{1}$H and $^{31}$P were not acquired from identical tissue volumes.
 
\textbf{ROI mismatch and $B_0$ inhomogeneity at 7~T.}

A voxel in the medial prefrontal cortex (mPFC) was selected \textit{a priori} given reports of altered prefrontal activity and metabolism in obesity and diabetes \cite{Li2023MolPsychiatry,Tataranni1999PNAS,Page2011JCI,Volkow2009Obesity,Le2006AJCN_DLPFCMealObesity}. The $^{1}$H SVS voxel was placed accordingly (Figure~\ref{fig:fig1}B, blue) and locally optimized using FASTMAP shimming, which improves $B_0$ homogeneity within a prescribed volume-of-interest \cite{Gruetter1993FASTMAP}. In anterior/frontal ROIs at 7~T, robust SVS generally benefits from careful $B_0$ optimization, with e.g., localized shimming and/or field-map-based shim updates\cite{Wilson2019Consensus,Juchem2020B0Shimming}. In the present interleaved setup, $B_0$ optimization for the $^{1}$H voxel relied on FASTMAP, as the single-channel double-tuned coil configuration did not support the GRE-based shimming workflow commonly used in dedicated $^{1}$H protocols.
 
For $^{31}$P, we chose an ultra-short TE whole-brain MRSI sequence (PETALUTE) rather than $^{31}$P SVS because chemical-shift dispersion and linewidth requirements are generally less stringent for $^{31}$P quantification than for $^{1}$H, and whole-brain coverage enables retrospective ROI definition and greater tolerance to regional variability in spectral quality \cite{Alcicek2025PETALUTE}. In practice, ROI-specific spectral quality in the anterior brain was insufficient, requiring a post hoc adjustment of the $^{31}$P ROI. The single-channel double-tuned volume coil exhibits a centrally peaked sensitivity profile, yielding higher $^{31}$P SNR toward the brain center and reduced sensitivity toward peripheral/anterior regions.  In addition, this limitation is consistent with the pronounced susceptibility-driven $B_0$ inhomogeneity in prefrontal regions at 7~T, particularly near air/tissue interfaces such as the paranasal sinuses, where vendor global shimming does not reliably provide optimal homogeneity across the full MRSI FOV\cite{Wilson2019Consensus,Juchem2020B0Shimming}.  Together, these factors reduced the robustness of anterior $^{31}$P spectra, and spectra were therefore extracted from a posterior ROI with more favorable coil sensitivity and field homogeneity (Figure \ref{fig:fig1}B). Accordingly, dedicated higher-sensitivity $^{31}$P coil configurations and advanced shimming (higher-order, dynamic and/or target-volume-specific) could improve anterior $^{31}$P PETALUTE spectral acquisition in future implementations \cite{Juchem2020B0Shimming,Alcicek2025PETALUTE}.

The resulting spatial mismatch limits direct statements about local coupling between $^{1}$H Glc+Tau and $^{31}$P energetics and instead only allows interpretation at the level of systemic modulation. Future implementations could mitigate this limitation by harmonizing ROIs across nuclei where feasible, and using advanced shimming strategies as higher-order, dynamic and/or target-volume-specific shimming, to improve $B_0$ homogeneity in anterior brain regions for both SVS and MRSI \cite{Juchem2020B0Shimming,Wilson2019Consensus}.

\textbf{Long protocol duration and motion/spectral drift.} The total experiment time ($\sim$120~min plus pre-scans) is a known limitation of the experiment setup. The long scan time, combined with infusion procedures and repeated blood-glucose sampling increases vulnerability to participant discomfort. In addition, subtle head motion and scanner-related frequency/drift effects, all together can lead to progressive degradation in spectral quality \cite{Xiao2026Thesis,Wilson2019Consensus,Juchem2020B0Shimming}. This is particularly relevant for repeated MRS blocks, where small head shifts can alter coil loading and ROI partial volume, and for repeated frontal SVS acquisitions where susceptibility-related $B_0$ instability and motion sensitivity are more pronounced in a volume-coil setting. Shortening the overall protocol, improving comfort and head stabilization, and incorporating prospective motion correction and/or dynamic shim updating would likely improve robustness \cite{Juchem2020B0Shimming}. For future studies, we suggest reducing baseline sampling to two blocks and, if a single steady-state clamp-stage is sufficient, limiting acquisition to Hyperglycemia~I (two interleaved blocks), which would reduce total scan time to approximately 71~min.

\textbf{Practical constraints of interleaved multinuclear clamp experiments.} Sustained blood-glucose clamping over long scan sessions is technically demanding, and infusion-line issues, as e.g. a clocked line, can compromise clamp-stage completeness and statistical power, as reflected by missing hyperglycemia blocks in one participant. Interleaving adds further constraints: SAR limits required participant-specific TR adaptations for the $^{1}$H STEAM blocks, leading to non-uniform temporal sampling across subjects and clamp-stages. Together, clamp interruptions and SAR-driven timing variability motivate the use of statistical frameworks that tolerate unbalanced designs, in this study linear mixed-effects models, but they also reduce sensitivity for clamp-stage comparisons. Future studies could improve robustness by adopting a conservative $^{1}$H TR from the outset e.g., matching the longest TR used here, $\sim$6~s, to standardize sampling across participants, and by incorporating redundant clamp monitoring, multichannel $^{31}$P coils and hardware improvements to reduce the likelihood of infusion-line failure.
The cohort size was small, and therefore inference relied primarily on within-subject repeated measures rather than between-subject variation. This increases sensitivity to outliers and missing blocks and limits stratified analyses (e.g., by sex, BMI, or HbA1c). The controlled glucose infusion and interleaved acquisition design nevertheless supports within-subject contrasts across glycemic stages, reducing confounding from inter-individual differences. Finally, ratio-based endpoints (PCr/Pi, $\gamma$ATP/Pi, and baseline-normalized Glc+Tau) improve stability but do not distinguish whether observed changes arise predominantly from the numerator, the denominator, or both.

\section{Conclusions}\label{sec5}
In summary, interleaved $^{1}$H/$^{31}$P acquisitions during acute hyperglycemia generated by continuous IV glucose infusion provided a robust $^{1}$H readout that tracked blood glucose, alongside smaller but significant changes in $^{31}$P ratios consistent with glycemia-linked modulation of high-energy phosphate homeostasis. Although the spatial mismatch between $^{1}$H and $^{31}$P sampling and the demands of a long session limited strictly local mechanistic interpretation, the combined measurements demonstrate the feasibility of capturing coordinated substrate- and energetics-related signatures within a single 7~T experiment.

Despite these constraints, the study provides a practical template for combining rapid whole-brain $^{31}$P MRSI with repeated $^{1}$H SVS during a glycemic intervention at 7~T.  Next steps are to to harmonize spatial sampling by improving frontal $^{31}$P quality, choosing a shared ROI and reduce experiment duration and shortening/unifying the sampling protocol.  With these refinements, interleaved multinuclear MRS could become a useful tool to probe acute metabolic flexibility and energetic regulation in humans under controlled systemic perturbations, including altered metabolic states such as obesity and diabetes.

\section*{Author contributions}
JJH, UEE, MB and MW conceived, designed, performed the experiments, and wrote the paper

\section*{Acknowledgments}
The authors gratefully acknowledge the nurses of the UNC Clinical Translational Research Unit including Amy Ellis and Janette Goins as well as Joseph Palmiotto's invaluable assistance with participant recruitment. Scanners were operated by local radiographers Amber Abernethy Leinwand, Robert Stewart Little, and Brittany Rachelle Weir.
\section*{Financial disclosure}

No Financial disclosure.

\section*{Conflict of interest}

The authors declare no potential conflict of interests.

\section*{Supporting information}

Grant support: R01DK123227 (JJH), P30124723 (NC Diabetes Research Center), UL1TR001111, The Obesity Society Early Career Research Grant (BCM), Wellcome Trust Collaborative Award 223131/Z/21/Z (UE).


\newpage
\section*{Supplementary Information}
\setcounter{figure}{0}
\renewcommand{\thefigure}{S\arabic{figure}}
\subsection*{Supplementary Figure S1: Z-scored time courses of blood glucose and spectroscopic measures}
\begin{figure}[h]
\center
\includegraphics[width=0.65\textwidth]{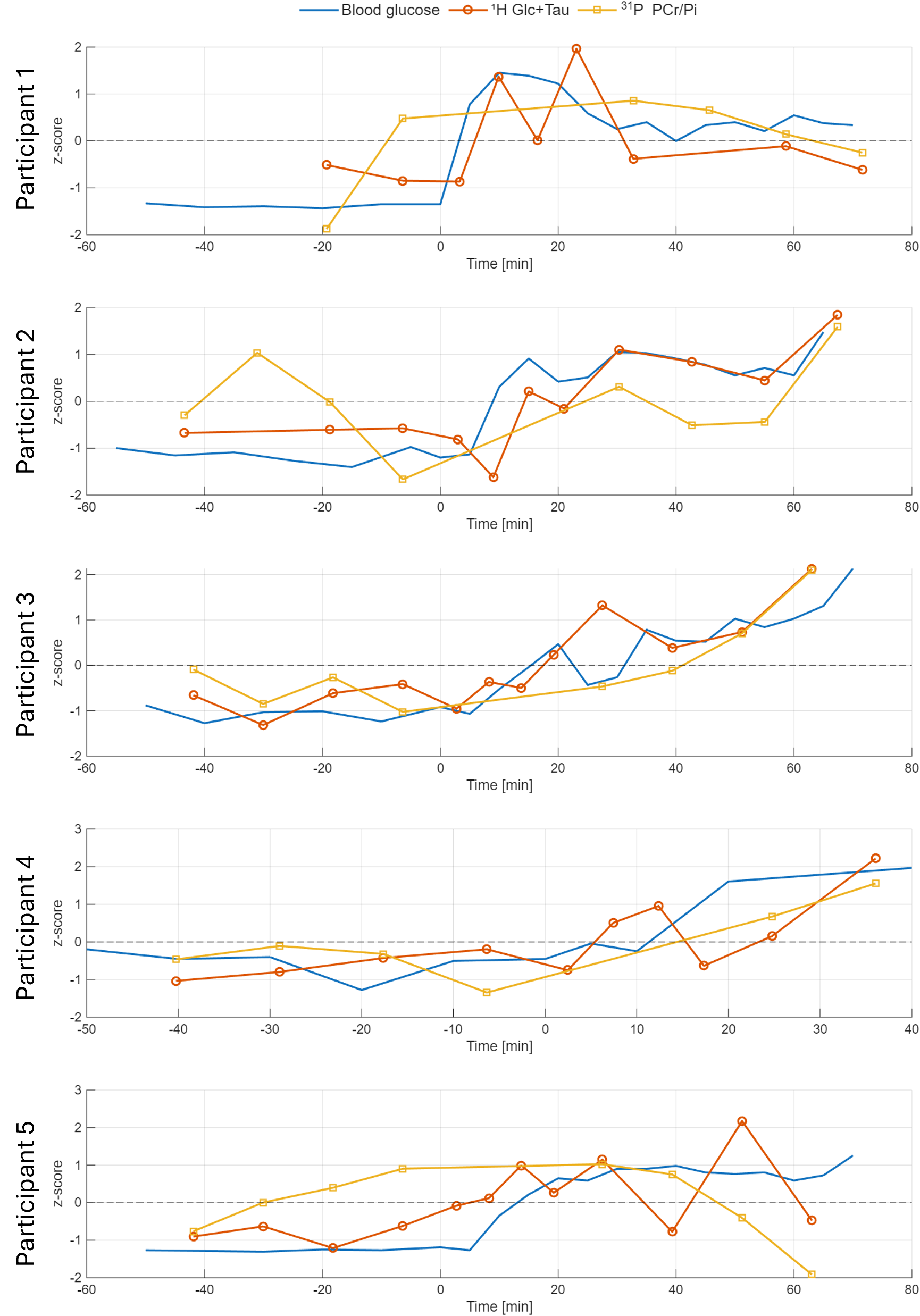}
\end{figure}
Supplementary Figure~S1 shows participant-wise time courses of blood glucose, $^{1}$H Glc+Tau, and $^{31}$P PCr/Pi expressed as $z$-scores (within-participant standardization) to facilitate visual comparison of temporal co-variation across modalities despite different units and dynamic ranges. This representation highlights that the rise in Glc+Tau largely coincides with increases in blood glucose, while PCr/Pi exhibits smaller and more heterogeneous temporal responses across participants. The $z$-score visualization complements the baseline-normalized trajectories presented in Figure~2 by emphasizing relative deviations from each participant’s mean rather than changes relative to the baseline reference.
\newpage
\subsection*{Supplementary Figure S2: Associations for $^{31}$P $\gamma$ATP/Pi}
\begin{figure}[h]
\center
\includegraphics[width=\textwidth]{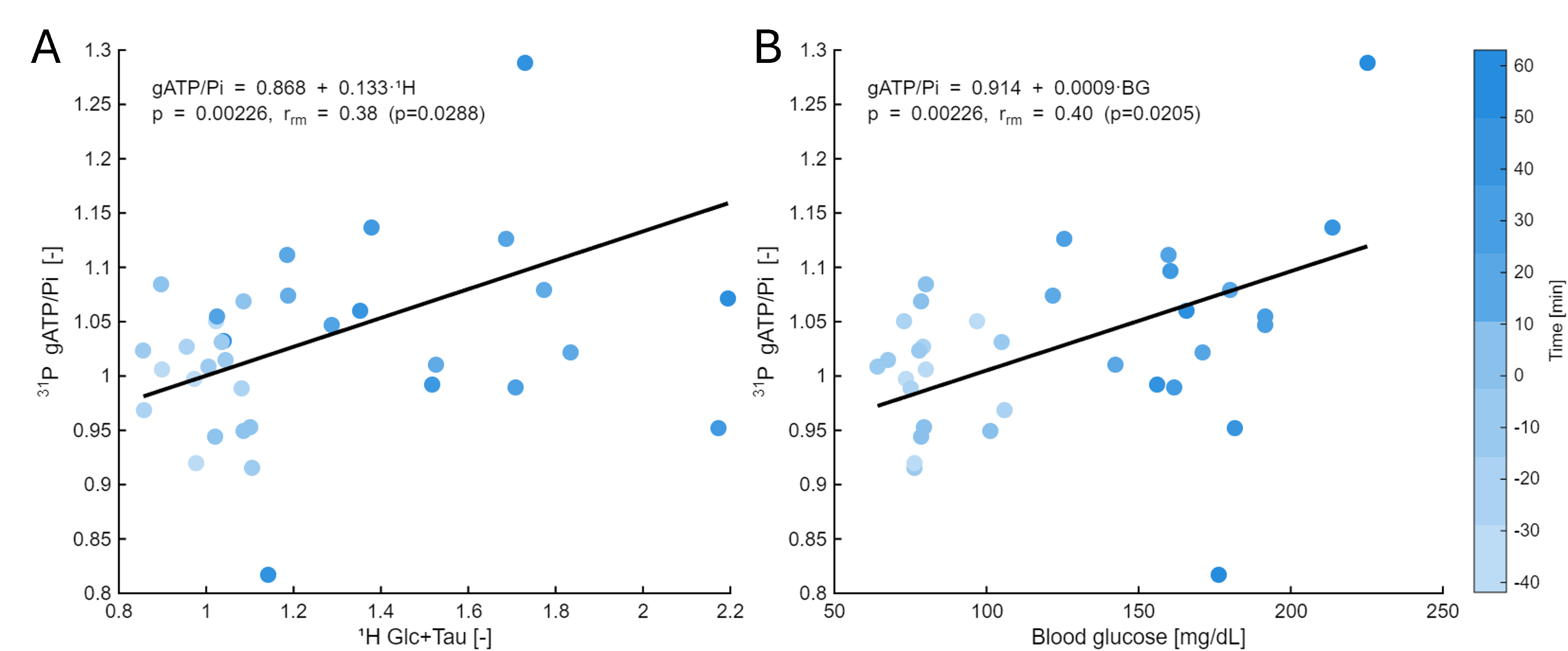}
\end{figure}
Supplementary Figure~S2 shows associations between $^{31}$P $\gamma$ATP/Pi and (A) $^{1}$H Glc+Tau as well as (B) blood glucose across participants and time points. Linear fits and repeated-measures correlation statistics are shown in-panel, with point color encoding time relative to infusion start. Consistent with the PCr/Pi analysis (Figure~3), $\gamma$ATP/Pi shows a modest but significant positive association with both the $^{1}$H glucose-related metric and systemic glycemia, supporting sensitivity of PETALUTE-derived energetic ratios to glycemia-linked modulation.

\bibliography{library}

\bibliographystyle{abbrv}

\end{document}